\documentclass[aps,pre,twocolumn,float]{revtex4}
\usepackage{amssymb,amsmath,latexsym}
\usepackage{epsfig}
\def\Fbox#1{\vskip1ex\hbox to 8.5cm{\hfil\fboxsep0.3cm\fbox{%
  \parbox{8.0cm}{#1}}\hfil}\vskip1ex\noindent}  

\begin{document}

\title{Finite Size Scaling for the Glass Transition: the Role of a {\em Static} Length Scale}
\author{Smarajit Karmakar$^{1}$ and Itamar Procaccia$^{2}$}
\affiliation{$^{1}$ Departimento di Fisica, Universit\'a di Roma ``La Sapienza '', Piazzale Aldo Moro 2, 00185, Roma, Italy,\\
$^{2}$ Dept. of Chemical Physics, The Weizmann Institute of Sceince, Rehovot 76100, Israel}
\date{\today}
\begin{abstract}
Over the last decade computer simulations have had an increasing role in shedding light on difficult statistical physical phenomena and in particular on the ubiquitous problem of the glass transition. Here in a wide variety of materials the viscosity of a super-cooled liquid increases by many orders of magnitude upon
decreasing the temperature over a modest range. A natural concern in these computer simulation  is the very small size
of the simulated systems compared to experimental ones, raising the issue of how to assess the thermodynamic limit.
Here we offer a theory for the glass transition based on finite size scaling, a method that was found very useful in the context of critical phenomena and other interesting problems. As is known, the construction of such a theory rests crucially on the existence of a growing {\em static} length scale upon decreasing the temperature. We demonstrate that the static
length scale that was discovered in Ref. \cite{12KLP} fits the bill extremely well, allowing us to provide a finite size
scaling theory for the $\alpha$ relaxation time of the glass transition, including predictions for the thermodynamic limit
based on simulations in small systems.
\end{abstract}
\maketitle

\section{Introduction}
Glasses are ubiquitous in nature; the problem of the glass transition where the viscosity of a super-cooled liquid increases by many orders of magnitude upon a temperature decrease of a couple of hundred degrees remains one of the major challenges in modern condensed matter physics. With the advent of powerful computers, numerical simulations played a vital role in revealing some of the crucial characteristic associated with this phenomenon \cite{09Cav,11BB}. In spite of this progress, due to still present limitations in computer power, simulations are still done in small sample sizes from thousands to a few tens of thousands of particles. Compared to experimental samples this is minute. Thus finite
size effects in the results are always a major concern, especially when one deals with a system where a characteristic length scale grows with decreasing temperature \cite{12KLP}.

Finite size effects are not always unwanted; in fact one can extract useful insights by studying these effects
carefully \cite{90B}. As an example the temperature dependence of the static length scale can be obtained from the system size 
dependence of some relevant observable by analyzing the data using finite size scaling. In the glass community the existence and 
usefulness of a static length scale is still debated; even if one can extract some static length scale by doing some sophisticated 
analysis \cite{03Ber, 08BBCGV, 09KDS},
the slow growth of this length scale makes it difficult to extract reliable insights . Nevertheless
finite size effects in the dynamics of the supercooled liquids certainly exist, notwithstanding the fact that the
characteristic length scale may not change very much. A careful analysis of these effects can shed a substantial light 
on the physics of the glass transition
\cite{91DIRP, 98DDPPG, 00KY, 00FP, 01BKB, 02DFPG, 03Ber,09KDS, 10KDS, 10KDSa, 10RRH, 96Horbach, 99BH, 03DH}. 

In \cite{12KLP}, we already showed that our proposed length scale is the same object which governs the finite size effects
exhibited in configurational entropy $S_c$ in the Kob-Andersen Model in 3 dimensions \cite{09KDS}. Further more $\tau_{\alpha}$ and $S_c$ are 
related via the empirical Adam-Gibbs relation \cite{AG} or may be via more sophisticated RFOT Theory \cite{RFOT, RFOT1, RFOT2}. 
So our length scale directly relates dynamics with thermodynamics in glass forming liquids via the Adam-Gibbs relation. Unfortunately 
we can not determine the exponents predicted in the RFOT Theory since the range of increase of our length scale is not sufficient to 
determine reliably the exponents.

In this study our main aim is to establish that our proposed length scale is a crucial object for the theory of glass transition
by demonstrating that it faithfully explains the finite size effects seen in relaxation times. In a similar attempt in \cite{10RRH} to 
explain the finite size effect seen in the relaxation time, it was argued that a 
large system of particles can be thought of as a collection of smaller subsystems called ``elementary units" which are coupled to each other 
by coupling constants. The finite size effects seen in the relaxation times can be thought of as a organized rearrangements of 
these elementary units. It is quite possible that our proposed length scale actually measures the size of these elementary units, but 
further study is needed to pinpoint this connection. It is also worth mentioning that in \cite{10RRH} a linear relationship has been 
proposed between the relaxation time and the system size for small system sizes. In contrast we propose a scaling function associated 
with the underlying growing static length scale.

A clear advantage of our proposed length scale \cite{12KLP} is that it is relatively easy to extract both in simulations and 
in experiments \cite{10GCSKB} as compared to the other length scales like point-to-set \cite{08BBCGV} and patch length scale \cite{11KL,11SL}. 
While we can compute our length scale for larger range of temperature this is prohibitive for the other length scales, thus at this time
there is not enough data to provide a conclusive comparison between these length scales.   

Below we present a systematic study of the system size dependence of the $\alpha$-relaxation time of different model glass forming systems 
in both two and three dimensions. We choose two completely different interaction pair potential- one with pure repulsion and one with short 
range repulsion and long range attraction to stress the generality of the finite size effects in the dynamics of glass forming systems 
\cite{12BBCKT}.

\section{Models and Methods}

Here we study two distinct models, one with pure power law repulsive interaction \cite{12KLP, 11KLPZ} and the second being
the well-studied Kob-Andersen Model \cite{95KA} in both 2 and 3 dimensions. We will denote the Kob-Andersen Model in 3-dimensions as 3d KA and
the slightly modified version of it for 2-dimensions as 2d mKA \cite{09BSPK}. The pure repulsive models will be referred to as 3d R10
and 2d R10 for 3-dimensions and 2-dimensions respectively. The usual Kob-Andersen model with a $80:20$ binary mixture in 2-dimension shows prominent clustering effects so the model was slightly modified and a $65:35$ binary mixture was used. This modified model shows no clustering behavior in all the simulated temperature range \cite{09BSPK, 12SKDS}. Throughout this
paper we use temperature units such that the Boltzmann constant equals unity.

The potential of the Kob-Andersen model is given by
\begin{equation}
V_{\alpha\beta}(r)=4\epsilon_{\alpha\beta}[(\frac{\sigma_{\alpha\beta}}{r})^{12}-
(\frac{\sigma_{\alpha\beta}}{r})^{6}]
\end{equation}
where $\alpha,\beta \in \{A,B\}$ and $\epsilon_{AA}=1.0$, $\epsilon_{AB}=1.5$,
$\epsilon_{BB}=0.5$, $\sigma_{AA}=1.0$, $\sigma_{AB}=0.80$, $\sigma_{BB}=0.88$. The
Interaction Potential was cut off at 2.50$\sigma_{\alpha\beta}$. We have performed the simulations
at six different temperatures in the range $T \in [{0.60,\, 0.43}]$ in 3-dimensions and $T \in [{0.90,\, 0.45}]$ in 2-dimensions at a number density $\rho = N/V = 1.20$.

The interaction potential for the pure repulsive model is given by
\begin{equation}\label{potential}
\phi\left(\!\frac{r_{ij}}{\lambda_{ij}}\!\right) =
\left\{ \begin{array}{ccl} \!\!\varepsilon\left[\left(\frac{\lambda_{ij}}{r_{ij}}\right)^{k} +
\displaystyle{\sum_{\ell=0}^{q}}c_{2\ell} \left(\frac{r_{ij}}{\lambda_{ij}}\right)^{2\ell}\right]
&\! , \! & \frac{r_{ij}}{\lambda_{ij}} \le x_c \\ 0 &\! , \! & \frac{r_{ij}}{\lambda_{ij}} > x_c
\end{array} \right.
\end{equation}
where $r_{ij}$ is the distance between particle $i$ and $j$, $\varepsilon$ is the energy scale, and $x_c$ is the
 dimensionless length for which the potential vanishes continuously with $q$ derivatives. The interaction
length scale $\lambda_{ij}$ between any two particles $i$ and $j$ is $\lambda_{ij} = 1.0\lambda$,
$\lambda_{ij} = 1.18\lambda$ and $\lambda_{ij} = 1.4\lambda$ for two `small' particles, one `large' and one `small'
particle and two `large' particle respectively. The coefficients $c_{2\ell}$ are given by
\begin{equation}
c_{2\ell} = \frac{(-1)^{\ell+1}}{(2q-2\ell)!!(2\ell)!!}\frac{(k+2q)!!}{(k-2)!!(k+2\ell)}x_c^{-(k+2\ell)}.
\end{equation}
We chose the parameters $x_c = 1.3854$, $k=10$ and $q=2$. The unit of length $\lambda$ is set to be the interaction length scale
of two small particles, and $\varepsilon$ is the unit of energy. We have performed the simulations at six different temperatures
in range $T \in [{0.80,\, 0.52}]$ at number density $\rho = 0.81$ in 3-dimensions and $T \in [{0.65,\, 0.48}]$ at $\rho=0.85$
for 2-dimensions.

We have simulated these  models in the $NVT$ ensemble using a Berendsen thermostat \cite{84BPGDH} to maintain the temperature. 
The system size was in the range
$N\in [128, 20164]$ and we used as many as $20$ to $50$ different samples to average the data depending on the system size. Generally
for small system sizes and at low temperature the fluctuations increase, demanding more extensive averaging of the
data to get reliable estimates of the wanted observables. For each system size and each temperature we measured
the $\alpha$-relaxation time and the characteristic length scale. The former is calculated from the decay of the overlap correlation function defined as
\begin{equation}
Q(t) = \frac{1}{N}\sum_{i=1}^{N} w(|\vec{r}_i(t) - \vec{r}_i(0)|),
\end{equation}
where the weight function $w(x) = 1$ if $x<0.30$ and zero otherwise. The $\alpha$-relaxation time $\tau_\alpha$ is the time where this correlation function goes to $1/e$ of its initial value i.e. $Q(\tau_\alpha) = 1/e$. The characteristic length scale was measured using the method proposed in Ref. \cite{12KLP}. The reader is encouraged to examine this reference for the details
of the method. In brief, it is based on the insight that the minimum eigenvalue of the Hessian matrix belongs to either
an elastic mode or to a plastic mode, depending on the system size. The characteristic length is the crossover
system size between these two possibilities. The Hessian matrix is obtained for every supercooled system at
temperature $T$ by employing the conjugate gradient minimization methods
to get an inherent structure (at temperature $T=0$), and then using the Lanczos method \cite{lancAlgo} to get the minimum eigenvalue.
\begin{figure*}
\includegraphics[scale = 0.50]{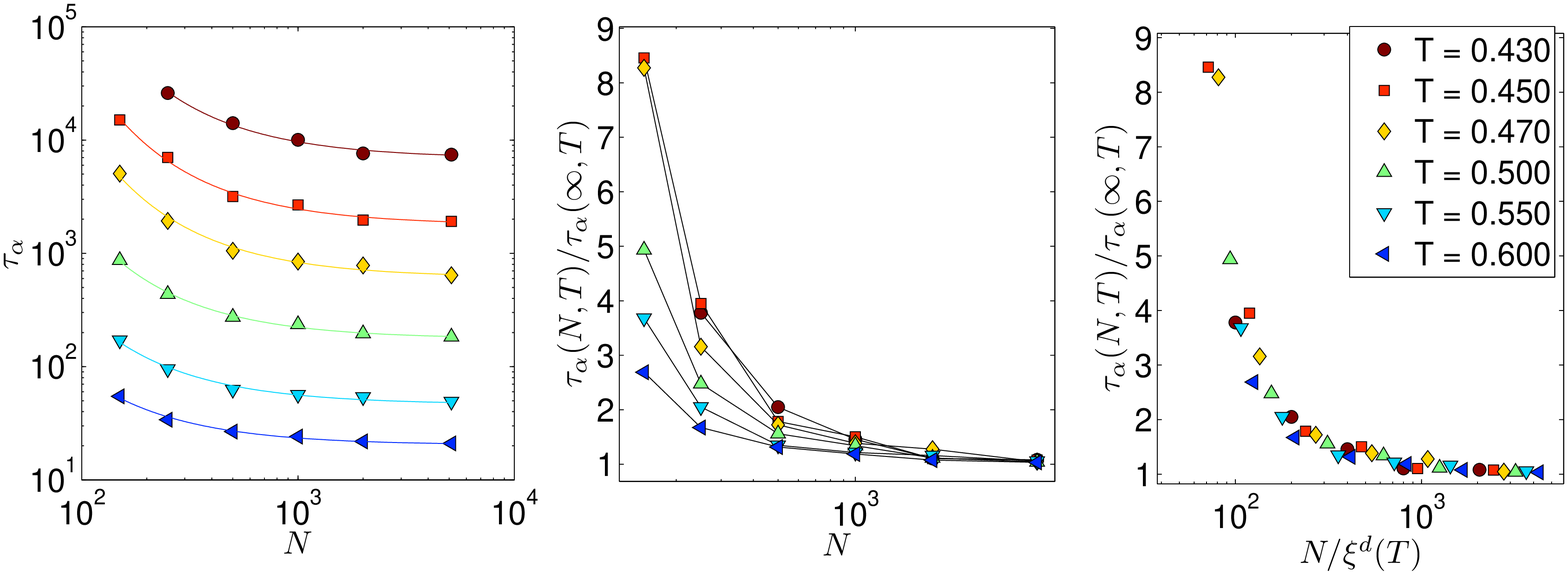}
\caption{(Color online) 3d KA Model : Left Panel: The system size dependence of the $\alpha$-relaxation time for the Kob-Andersen Model in 3-dimensions for different
temperatures. The lines are fits to the data using the fitting function $f(x) = A + B/x$ to extract the large $N$ value
$\tau_{\alpha}(N\to\infty,T)$ of the relaxation time. Middle Panel: The y-axis is rescaled by this value such that all the data for large $N$ converge to unity.
Note that this procedure does not collapse the data indicating the existence of a length scale. Right Panel: The
complete collapse of the data using the length scale obtained in \cite{12KLP} (see text for details).}
\label{ka3d}
\end{figure*}

\section{Results}
\begin{figure*}
\includegraphics[scale = 0.46]{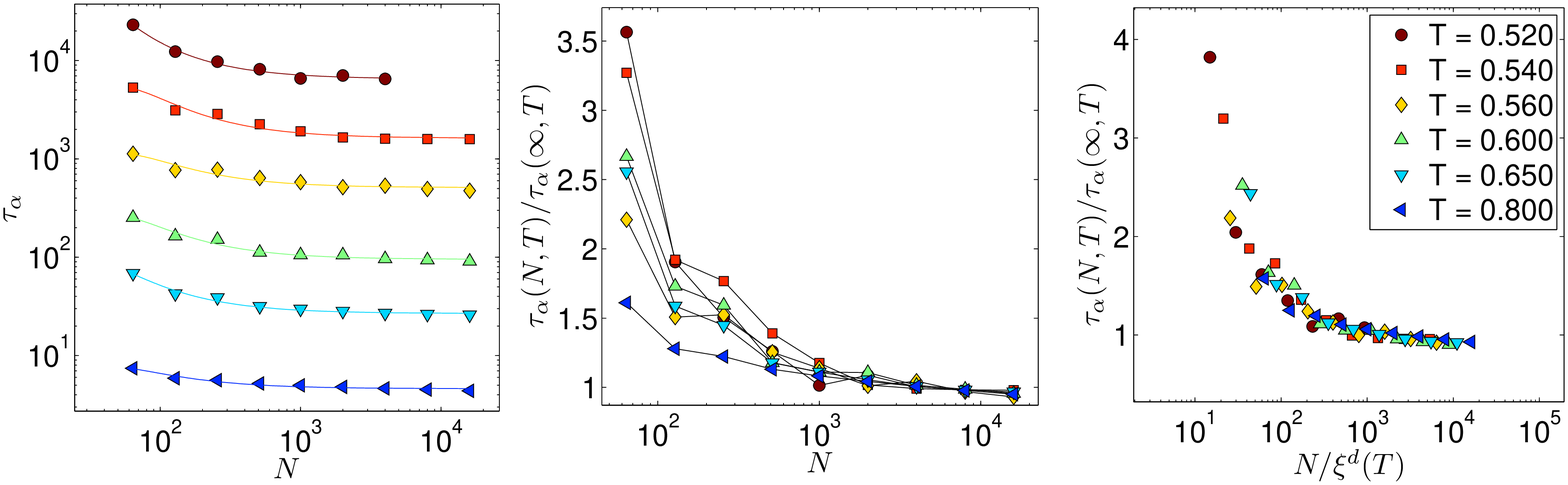}
\caption{(Color online) 3d R10 Model: Left Panel: The system size dependence of the $\alpha$-relaxation time for a system interacting via a $1/r^{10}$
pure power law repulsive potential in 3-dimensions for different temperatures. The lines are fits to the data using the fitting function
$f(x) = a + b/x$ to extract the large $N$ value
$\tau_{\alpha}(N\to\infty,T)$ of the relaxation time. Middle Panel: same rescaling as in the previous figure. Right Panel: The complete collapse of the data using the length scale obtained in \cite{12KLP} (see text for details).}
\label{R103d}
\includegraphics[scale = 0.54]{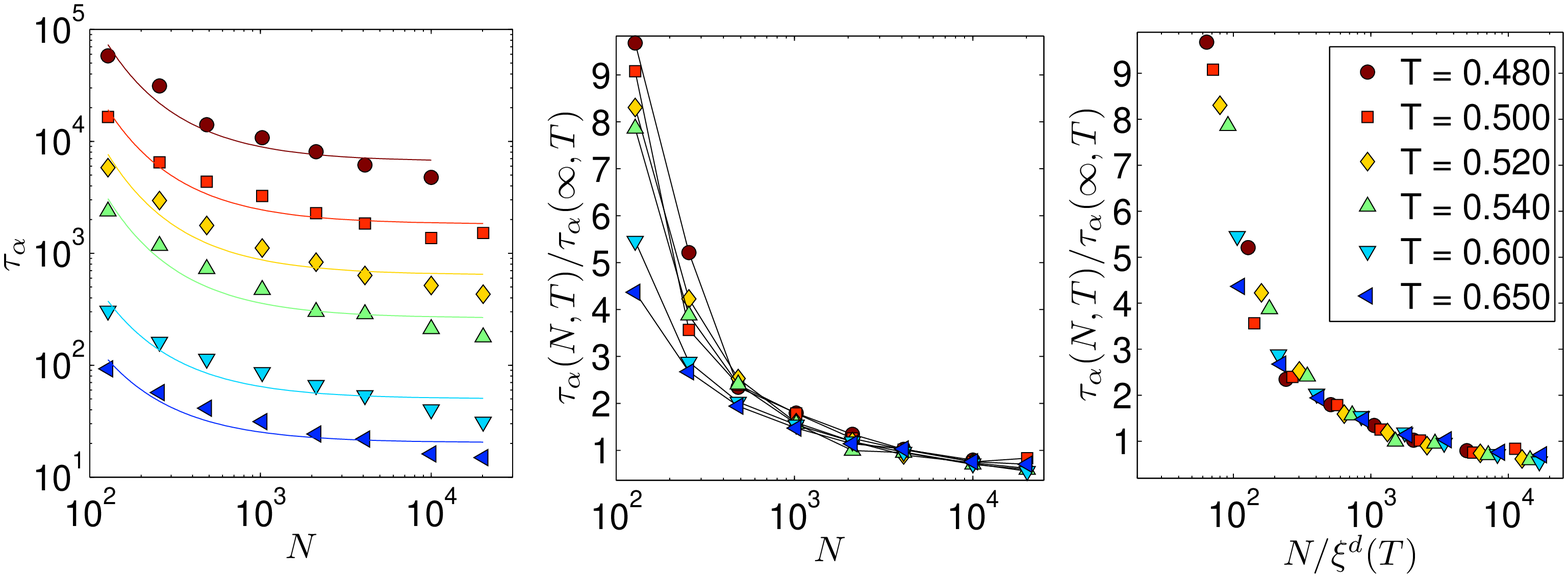}
\caption{(Color online) 2d R10 Model: Left Panel: The system size dependence of the $\alpha$-relaxation times for a system interacting via a $1/r^{10}$
pure power law repulsive potential 2-dimensions for different
temperatures. The middle and right panels repeat the procedure explained in Fig. 1}
\label{R102d}
\includegraphics[scale = 0.50]{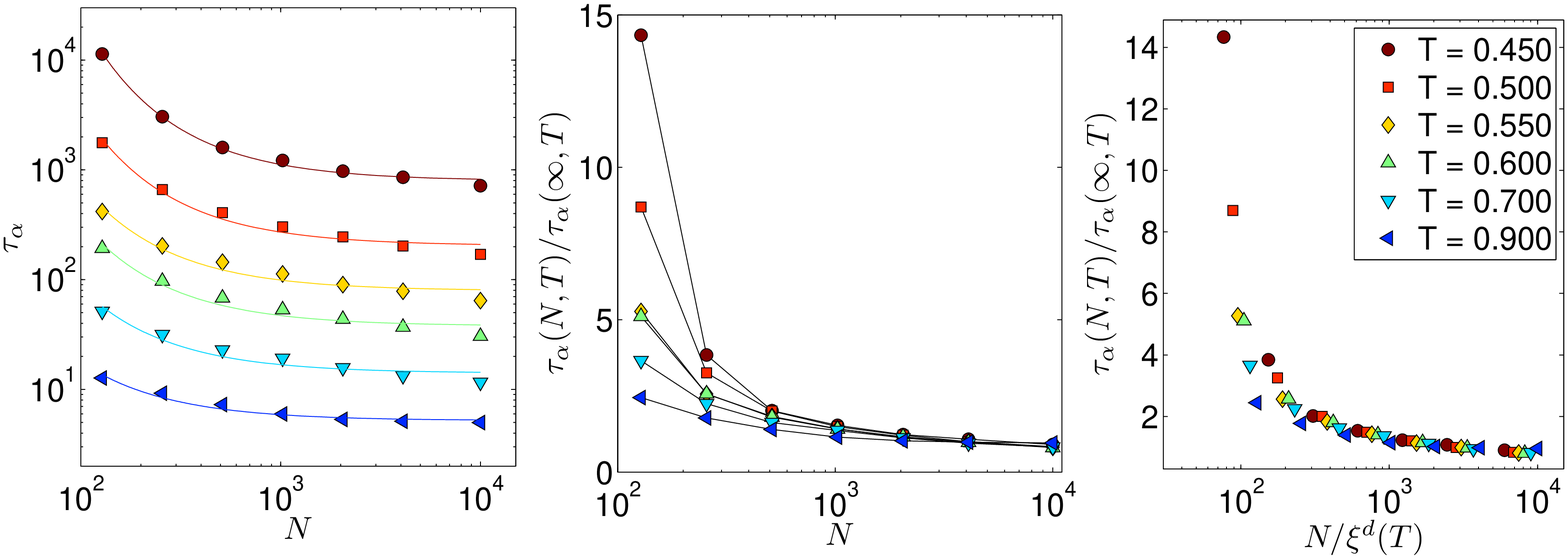}
\caption{(Color online) 2d mKA Model: Left Panel: The system size dependence of the $\alpha$-relaxation times for 
modified Kob-Andersen Model
(mKA Model, see text for details) in 2-dimensions for different
temperatures. The fits and the middle and right panels follow the procedure described in Fig. 1}
\label{mka2d}
\end{figure*}

In the left panel of Fig.\ref{ka3d}, we present the system size dependence of the $\alpha$-relaxation time $\tau_\alpha$
for the Kob-Andersen model in 3-dimensions, for different temperatures. The data indicates a substantial system size
dependence of the relaxation time which appears to increase slowly with decreasing the temperature. This is an indication
that the length scale which governs this behavior increases also with decreasing the temperature.
Physically we expect the relaxation time to decrease with
increasing system size if one asserts the existence of a cooperative length scale. This length scale becomes relevant 
to the system's dynamics when the temperature goes bellow the so-called
`onset temperature' \cite{01sastry} which separates the characteristic Arrhenius regime from the `fragile' regime
of the temperature dependence of the relaxation time, cf. \cite{12HKPZ}. The argument is as follows :
For a system of size $N$ at temperature $T$, the typical free energy barrier it needs to cross for relaxation is
\begin{equation}
\Delta F(N,T) = \Delta E(T) - T\log[g(N,T)],
\end{equation}
where $\Delta E(T)$ is the potential energy barrier. We expect the potential barrier to be determined by the energy landscape which in general does not depend on the
system size. On the other hand $\log[g(N,T)]$ is the entropic contribution to the free energy barrier. Now at temperatures lower than the
onset temperature \cite{01sastry} the cooperative relaxation mechanism plays the main role for relaxation and one expects the
degeneracy factor for this case to be system size dependent when the system size is small.

Consider a system of $N$ particles in a cubic box of size $L$ at some temperature $T$. In the non-Arrhenius regime with a typical length $\xi(T)$, we expect for a system with $L << \xi(T)$, that any relaxation event will occur on the scale $L$ in order to be successful. So the degeneracy factor will grow with system size. For a system with $L >> \xi(T)$ we expect $g(N,T) \sim const$ without any change when the system size grows. Thus the free energy barrier can be represented as a scaling function of $L/\xi(T)$ according to
\begin{equation}
\Delta F(N,T) = \Delta E(T) - T\log\left[A f\left( \frac{L}{\xi(T)}\right)\right],
\end{equation}
where $A$ is a proportionality constant independent of the temperature, and obviously
\begin{equation}
f(x)\to Const(T) \quad \text {when}~x\to \infty \ . \label{fofx}
\end{equation}
Next we calculate the system size dependence of the relaxation time for different temperatures, using the estimate
 $\tau_\alpha\approx \tau_0 \exp[\Delta F(N,T)/T]$ where $\tau_0$ is a typical attempt time, cf. \cite{12HKPZ}. The result is again
 a scaling function of $L/\xi(T)$, reading
\begin{eqnarray}
\tau_{\alpha}(N,T) &\sim& \tau_0 \exp\left( \frac{\Delta F(N,T)}{T} \right) \sim \tau_0 \frac{\exp\left( \frac{\Delta E(T)}{T} \right)}
{A f\left( \frac{L}{\xi(T)}\right)}\nonumber\\
&\sim& \tau_0 \exp\left( \frac{\Delta E(T)}{T} \right){\cal F}\left( \frac{L}{\xi(T)}\right).
\label{tauScalingAnsatz}
\end{eqnarray}

From the property (\ref{fofx}) of the scaling function we confirm that $\tau(N,T)\to Const(T)$ when $N\to \infty$.

Examine the consequences of these arguments for the data of the Kob-Andersen model, cf Fig. \ref{ka3d}. In the left
panel we present the $\alpha$ relaxation time for different temperatures and systems sizes.
In the middle panel of Fig.\ref{ka3d} we rescaled the y-axis by the large $N$ value of of the relaxation time $\tau_\alpha(\infty,T)$ 
for the different temperatures. We estimated this value by fitting the data with a functional form $f(x) = a + b/x$ ; One sees that 
this fitting is adequate for obtaining the large $N$ value of the relaxation time. In the right panel of Fig.\ref{ka3d}, we present 
the full data collapse which is obtained by rescaling the system size using the length scale obtained in \cite{12KLP}. We stress that
no adjustments were necessary and none were made to these length scales, they were employed as obtained. Now one may ask whether 
a similar collapse can be done using the other static length scales mentioned above. In a recent study 
\cite{12HMR} it was shown that the point-to-set length scale in the Kob-Andersen model in three dimension changes by a factor of $2.2$ in
the temperature range $T\in [1.00,0.550]$. Assuming that we can extrapolate these data to lower temperature we find  
point-to-set length scale changes by similar factor in the temperature range of our study. Thus it appears possible to collapse 
the relaxation time data using the point-to-set length scale in this temperature range. This suggests that point-to-set length 
scale is proportional to our length scale. It remains a future work to compute point-to-set length scale to lower temperature to 
confirm this connection.

\begin{figure}
\includegraphics[scale = 0.40]{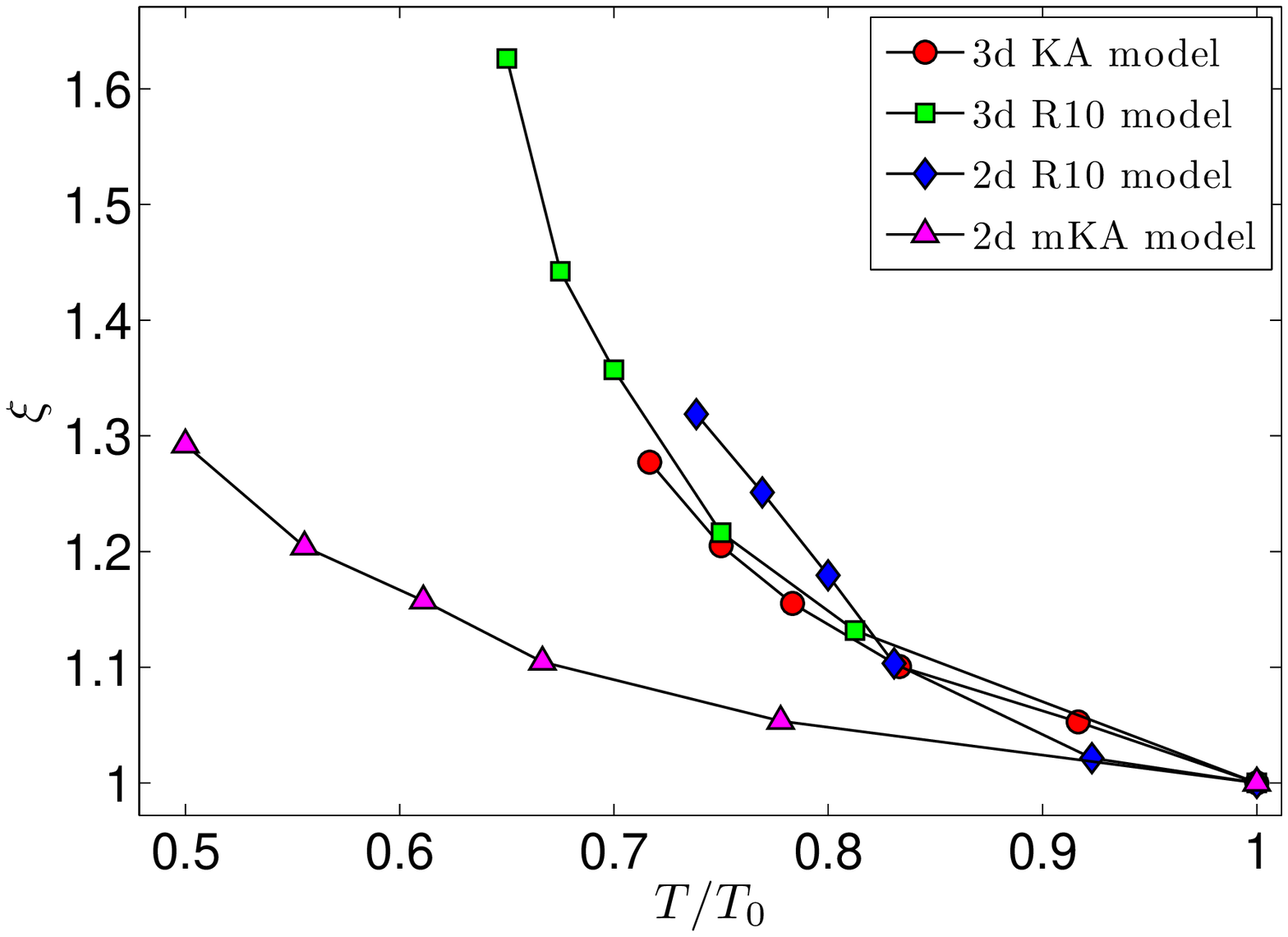}
\caption{(Color online) Temperature dependence of the static length scale $\xi$ for different models studied.}
\label{xiT}
\includegraphics[scale = 0.39]{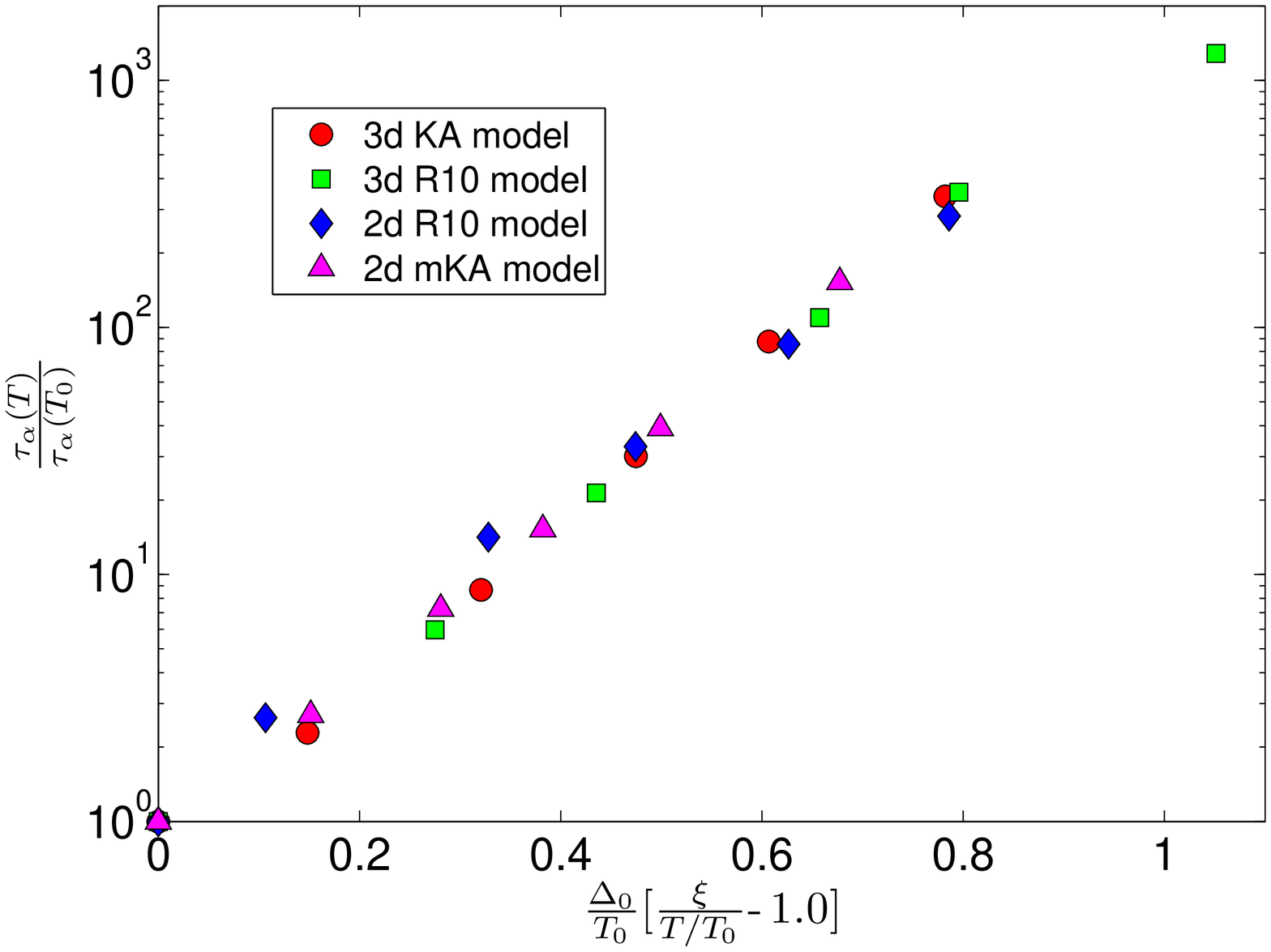}
\caption{(Color online) Relaxation time in the thermodynamic limit  plotted according to Eq. \ref{ansatz} ( see
text for details) for all the four studied models. The apparent universality is encouraging. The parameter $\Delta_0$ used
here (see text for details) is model dependent and numbers are quoted in the text.}
\label{tauXi}
\end{figure}

In Fig.\ref{R103d} we present the results of a similar analysis for the model system with pure repulsive interaction (3d R10) in 3-dimensions.
The data collapse observed in the right panel is again quite good. We present similar analysis
in 2-dimensions for the model with pure repulsive interaction in Fig.\ref{R102d} and for the modified Kob-Andersen model in
Fig.\ref{mka2d}. In Fig.\ref{xiT}, we plotted the temperature dependence of the correlation length for all the models studied to compare
the model to model variation of the growth of this length scale with decreasing temperature.

\section{Universal Relation Between $\tau_{\alpha}$ and $\xi$}
Having found that the typical scale fits the bill for finite size scaling, we should ask next whether there is 
a direct, and possibly universal relation, between the length scale and the time scale.
We start with the following ansatz for the relation between structural relaxation time $\tau_{\alpha}$ in the thermodynamic limit 
and the static correlation length $\xi$.
\begin{equation}
\tau_{\alpha}(T) \propto \exp\left[ \frac{\Delta_0 \xi(T)^{\psi}}{T} \right],
\end{equation}
where $\Delta_0$ is a non-universal coefficient that depends on the details of the glass former. To set a scale to this relation we choose a 
reference temperature $T = T_0$ where the typical length is $\xi_0 = 1.0$. Then the relaxation time at that temperature is
\begin{equation}
\tau_\alpha(T_0) \propto \exp\left[ \frac{\Delta_0}{T_0}\right],
\end{equation}
So we have the following relation
\begin{eqnarray}
\log\left[ \frac{\tau_\alpha(T)}{\tau_\alpha(T_0)} \right] &=& \frac{\Delta_0 \xi^{\psi}}{T} - \frac{\Delta_0}{T_0}\nonumber\\
&=&  \frac{\Delta_0 }{T_0} \left[ {\xi}^{\psi}/\frac{T}{T_0} - 1.0 \right]\ .
\label{ansatz}
\end{eqnarray}
We reiterate that in the above equation the pre-factor $s \equiv \frac{\Delta_0}{T_0}$ is not known apriori for different models. 
In Fig.\ref{tauXi} we plotted the relaxation time data according to Eq.\ref{ansatz}, with $\Delta_0 = 1.0, 0.70, 1.0$,and $0.4274$ for KA 
model in 3D, R10 model in 3D, R10 model in 2D and modified KA model in 2D respectively. We choose $\psi = 1.0$ for all the models. 
The collapse of the data indicates an encouraging possible universal relation between the static length scale and the relaxation time.

To conclude, we have studied the finite size effects in the dynamics of supercooled liquids for different models systems in both 2 and 3
dimensions. In this article we have focused on the system size dependence of the $\alpha$-relaxation time and found that the
finite size effects can be very well explained by the static length scale that we proposed in \cite{12KLP} for all these different models.
In a recent study \cite{12BBCKT}, it was argued that a growing static length scale is necessary to explain the finite size effects
seen in the simulational studies of model glass forming liquids. We want to stress that our proposed length scale \cite{12KLP} is indeed
{\em that} length scale which faithfully explains all the finite size effects seen in the simulation results. 

Further more we realized that 
there is a possibility of universal relation between this length scale and the $\alpha$-relaxation time when relaxation time for
different models are plotted as a function of the length scales, all the data fall nicely on a master curve as also recently 
reported in \cite{12HMR}. This connection again seems to suggest that point-to-set length scale might be proportional 
to our proposed length scale. We reiterate that further study is needed to confirm this relation and care should be taken as these 
length scales and the relaxation times are not varying  by much in the temperature range for which we could do the simulations. 
We hope that our work will inspire other researchers to do further studies to delineate, especially in the low temperature range, 
the validity of this apparent universal behaviour between the relaxation time and the static length scale.

\acknowledgments
This work had been supported in part by an advanced ``ideas" grant of the European Research Council, the Israel Science Foundation and the German Israeli Foundation. We thank Shiladitya Sengupta, JNCASR, Bangalore for many useful discussions.


\begin{thebibliography}{99}
\bibitem{09Cav}
A. Cavagna Phys. Rep. {\bf 476}, 51 (2009).

\bibitem{11BB}
L. Berthier and G. Biroli, Rev. Mod. Phys. {\bf 83}, 587-645 (2011).

\bibitem{12KLP}
S. Karmakar, E. Lerner, I. Procaccia, Physica A , {\bf 391}, 1001 (2012)

\bibitem{90B}
K. Binder , Z Phys B {\bf 43}, 119 –140 (1990).

\bibitem{08BBCGV}
G. Biroli, J.-P. Bouchaud, A. Cavagna, T. S. Grigera,and P. Verrochio, Nature Phys. {\bf 4}, 771 (2008).

\bibitem{03Ber}
L. Berthier, Phys. Rev. Lett. {\bf 91}, 055701-1 (2003).

\bibitem{09KDS}
S. Karmakar, C. Dasgupta and S. Sastry, Proc. Natl. Acad. Sci. U.S.A. {\bf 106}, 3675 (2009).

\bibitem{91DIRP}
C. Dasgupta, A.V. Indrani, S. Ramaswamy, M.K. Phani, Europhys Lett {\bf 15}, 307–312 (1991).

\bibitem{98DDPPG}
C. Donati, J.F. Douglas, S.J. Plimpton, P.H. Poole, S.C. Glotzer, Phys Rev Lett {\bf 80}, 2338 (1998).

\bibitem{00KY}
K. Kim and R. Yamamoto, Phys Rev E {\bf 61}, R41 (2000).

\bibitem{00FP}
S. Franz, G. Parisi, J. Phys. Condens. Matter. {\bf 12}, 6335 (2000).

\bibitem{01BKB}
C. Brangian, W. Kob, K. Binder, Europhys Lett {\bf 53}, 756 (2001).

\bibitem{02DFPG}
C. Donati, F. Franz, G. Parisi, S.C. Glotzer, J. Non-Cryst. Solids. {\bf 307}, 215 (2002).

\bibitem{10KDS}
S. Karmakar, C. Dasgupta and S. Sastry, Phys. Rev. Lett. {\bf 105}, 015701 (2010).

\bibitem{10KDSa}
S. Karmakar, C. Dasgupta and S. Sastry, Phys. Rev. Lett. {\bf 105}, 019801 (2010).

\bibitem{10RRH}
C. Rehwald, O. Rubner, and A. Heuer, Phys. Rev. Lett. {\bf 105}, 117801 (2010).

\bibitem{96Horbach}
J. Horbach et al., Phys. Rev. E {\bf 54}, R5897 (1996).

\bibitem{99BH}
S. Buchner and A. Heuer, Phys. Rev. E {\bf 60}, 6507 (1999).

\bibitem{03DH}
D. Doliwa and A. Heuer, J. Phys.: Condens. Matter {\bf 15} S849 (2003).


\bibitem{AG}
G. Adams and J.H. Gibbs, J. Chem. Phys. {\bf 43}, 139 (1965).

\bibitem{RFOT}
T.R. Kirkpatrick, D. Thirumalai, P.G. Wolynes, Phys Rev A {\bf 40}, 1045 (1989).

\bibitem{RFOT1}
V. Lubchenko, P.G. Wolynes, Annu Rev Phys Chem {\bf 58}, 235 (2007).

\bibitem{RFOT2}
G. Biroli and J.-P. Bouchaud, {\it Structural Glasses and Supercooled Liquids: Theory, Experiment, 
and Applications} - edited by P.G. Wolynes, V. Lubchenko, John Wiley \& Sons, (2012).

\bibitem{10GCSKB}
A. Ghosh, V. K. Chikkari, P. Schall, J. Kurchan and D. Bonn
Phys. Rev. Lett., {\bf 104}, 248305 (2010).

\bibitem{11KL}
J. Kurchan and D. Levine, J. Phys. A: Math. Theor. {\bf 44}, 035001 (2011).

\bibitem{11SL}
F. Sausset and D. Levine, Phys. Rev. Lett. {\bf 107}, 045501 (2011).

\bibitem{12BBCKT}
L. Berthier, et. al  - arXiv.1203.3392 preprint 2012.

\bibitem{11KLPZ}
S. Karmakar, E. Lerner, I. Procaccia and J. Zylberg, Phys. Rev. E {\bf 83}, 046106 (2011).

\bibitem{95KA}
W. Kob and H. C. Andersen, Phys. Rev. E  {\bf 51}, 4626 (1995).

\bibitem{09BSPK}
R. Bruning, D.A. St-Onge, S. Patterson, W. Kob, J. Phys.: Condens. Matter {\bf 21} 035117, (2009).

\bibitem{12SKDS}
S. Sengupta, S. Karmakar, C. Dasgupta, and S. Sastry (unpublished).

\bibitem{84BPGDH}
H.J.C. Berendsen, J.P.M. Postma, W.F. van Gunsteren, A. Dinola, and J.R. Haak, J. Chem. Phys. {\bf 81}, 3684 (1984).

\bibitem{lancAlgo}
http://en.wikipedia.org/wiki/Lanczos\_algorithm.


\bibitem{12HKPZ}
H.G.E. Hentschel, S. Karmakar, I. Procaccia and J. Zylberg, - arXiv:1202.1127 preprint 2012.

\bibitem{01sastry}
S. Sastry, Nature {\bf 409}, 164 (2001).


\bibitem{12HMR}
G.M. Hocky, T.E. Markland, D.R. Reichman - Phys. Rev. Lett. {\bf 108}, 225506 (2012).

\end{thebibliography}
\end{document}